\documentclass[11pt]{article}
\usepackage{amsmath,amssymb,cite}
\usepackage{fullpage}
\usepackage{cleveref}
\usepackage{comment}

\newcommand{\ifabs}[2]{#1}
\renewcommand{\ifabs}[2]{#2}

\newcommand{\polylog}{\text{polylog }}

\newtheorem{theorem}{Theorem}

\newtheorem{definition}{Definition} 
\newtheorem{lemma}[theorem]{Lemma}

\newtheorem{corollary}[theorem]{Corollary}

\makeatletter
\input float.sty\relax
\makeatother

\newcommand{\Tabs}{12:\=11\=11\=11\=11\=11\=11\=11\kill}
\floatstyle{ruled}
\newfloat{alg}{htbp}{alg}
\floatname{alg}{Algorithm}
\newenvironment{algorithm}{%
      \begin{alg}%
      \begin{tabbing}\Tabs}%
      {\end{tabbing}\end{alg}}

\newcommand{\qedsymb}{\hfill{\rule{2mm}{2mm}}}
\newenvironment{proofof}[1]{\begin{trivlist}
\item[\hspace{\labelsep}{\bf\noindent Proof of #1: }]
}{\qedsymb\end{trivlist}}

\begin{document}

\title{Bounded-Contention Coding for Wireless Networks in the High SNR Regime\thanks{Massachusetts Institute of Technology,
Computer Science and Artificial Intelligence Laboratory and Research Laboratory of Electronics. ckeren@csail.mit.edu, haeupler@mit.edu, lynch@csail.mit.edu, medard@mit.edu.}}
%

%

\author{Keren Censor-Hillel 
\and Bernhard Haeupler
\and Nancy Lynch
\and Muriel M\'{e}dard}

\maketitle

\begin{abstract}
Efficient communication in wireless networks is typically challenged by the possibility of interference among several transmitting nodes. Much important research has been invested in decreasing the number of collisions in order to obtain faster algorithms for communication in such networks.

This paper proposes a novel approach for wireless communication, which embraces collisions rather than avoiding them, over an additive channel. It introduces a coding technique called \emph{Bounded-Contention Coding (BCC)} that allows collisions to be successfully decoded by the receiving nodes into the original transmissions and whose complexity depends on a bound on the contention among the transmitters.

\sloppy{BCC enables \emph{deterministic} local broadcast in a network with $n$ nodes and at most $a$ transmitters with information of $\ell$ bits each within $O(a\log{n}+a\ell)$ bits of communication with full-duplex radios, and $O((a\log{n}+a\ell)(\log{n}))$ bits, with high probability, with half-duplex radios. When combined with random linear network coding, BCC gives \emph{global} broadcast within $O((D+a+\log{n})(a\log{n}+\ell))$ bits, with high probability. This also holds in dynamic networks that can change arbitrarily over time by a worst-case adversary. When no bound on the contention is given, it is shown how to probabilistically estimate it and obtain global broadcast that is adaptive to the true contention in the network.}
\end{abstract}

\section{Introduction}
Handling interference in wireless networks is a fundamental challenge in designing algorithms for efficient communication. When two devices that are near each other transmit at the same time the result is a collided signal. In order to enable the receivers to obtain the original information that was sent, thereby achieving efficient communication that allows the design of fast algorithms for wireless networks, much important research has been invested in scheduling the transmissions in a way that avoids collisions as much as possible.

Avoiding collisions basically requires some type of symmetry breaking among the nodes that want to transmit, to prevent them from transmitting at the same time. Simple solutions like \emph{Time Division Multiple Access} (TDMA), which assigns predetermined slots according to node IDs, are expensive in situations where not all of the nodes want to transmit, since their costs depend on the total number of nodes rather than on the number of actual transmitters. One can improve this solution by allowing nodes that cannot interfere with each other to share a slot, so that the number of slots depends on the node degrees in the network graph rather than on the total number of nodes. However, this requires the nodes to have information regarding the topology of the network, and is still expensive in cases where the contention is less than the node degrees.
One successful approach for avoiding collisions in wireless networks is to allow each node to use a schedule of probabilities to decide whether to transmit at each time slot~\cite{BarYehudaGI87,KowalskiP2005,CzumajR2003}. These algorithms typically guarantee a high probability of successful transmissions after some bounded number of attempts.

In this paper we provide a coding framework for coping with collisions in a wireless communication model abstraction called the \emph{finite-field additive radio network model}, where a collision of transmissions optimally coded for an \emph{Additive White Gaussian Noise (AWGN)} channel with multiple-user interference is represented to be equivalent to the element-wise XOR of a string of bits representing the original transmissions. More generally,  collisions can be modelled as being equivalent to the sum, symbol-wise, of the elements of vectors over a finite field, where the transmission of a user is represented as a vector in that finite field, the XOR case being the special case where the field is $\mathbb{F}_2$. Such a model has been shown to be approximately valid in a high SNR (signal-to-noise ratio) regime, abstracting away the effect of noise in the channels and allowing us to concentrate on the theoretical aspects of the interference among transmissions. Such a model in effect replaces the traditional information-theoretic setting of Gaussian inputs for an AWGN channel with an approximate finite algebraic construct~\cite{KimEYM2011, AvestimehrDT2011}. Such additive models have been shown, albeit without a finite field construct, to be effective in high SNR settings even in the absence of underlying capacity-achieving codes, for instance in such systems as zig-zag decoding \cite{GollakotaK08}, which can be modelled algebraically \cite{ParandehGheibiS10}.  

In this additive model, our key observation is that \emph{only if not all messages are valid transmissions then a sum representing a collision might  indeed be uniquely decodable}. However, we do not wish to restrict the information the users may send in the wireless system. Instead, we propose encoding the information sent into restricted sets of signals that \emph{do} allow unique decoding when they collide. A node receiving a collision can then uniquely decode the signal to obtain the original unrestricted messages. Clearly, for information-theoretic reasons, we cannot hope to restrict the transmissions without the cost of some overhead in the amount of information sent. The challenge, then, is to find codes that allow unique decoding in the above setting with the shortest possible codewords. Under our high SNR assumption, we consider both half-duplex (sometimes termed time-division duplex - TDD) and full-duplex channels. While the TDD model is by far the most common current mode of operation, high SNR conditions can allow full-duplex operation.

\subsection{Our contributions}
\paragraph{The Bounded-Contention Coding (BCC) Framework:} 
We define a new class of codes, which are designed for settings in which the number of transmitters is bounded by a known constant $a$. Each such code consists of an individual code for each node, and has the desirable property that, when codewords from at most $a$ different transmitting nodes are summed up, the result can be uniquely decoded into the original transmissions. This decoding process does not require that the nodes know the identities of the transmitters. Moreover, the active nodes may change from round to round. We show simple constructions of Bounded-Contention Codes, where the length of the codewords depends on both the known contention bound and the total number of nodes, but the dependency on the total number of nodes is only logarithmic.

\paragraph{Distributed computation using BCC:} 
Using the new Bounded-Contention Coding technique, we show how to obtain local and global broadcast in both single-hop and multi-hop networks. BCC enables \emph{deterministic} local broadcast in a network with $n$ nodes and at most $a$ transmitters with information of $\ell$ bits each within $O(a\log{n}+a\ell)$ bits of communication with full-duplex radios, and $O((a\log{n}+a\ell)(\log{n}))$ bits, with high probability, with half-duplex radios. When combined with random linear network coding, BCC gives \emph{global} broadcast within $O((D+a+\log{n})(a\log{n}+\ell))$ bits. These results also hold in highly dynamic networks that can change arbitrarily over time under the control of a worst-case adversary. 

Further, we show how to remove the assumption that the nodes know a bound $a$ on the contention, by developing a method for handling unknown contention (or contention that varies over space and time), which is common in wireless networks. Also, while it may be reasonable to assume a bound on the contention, it is often the case that the \emph{actual} contention is much smaller.

\subsection{Related work}

The finite-field additive radio network model of communication considered in this paper, where collisions result in an addition, over a finite field, of the transmitted signals, was previously studied in~\cite{AvestimehrDT2011,KimEYM2011}, where the main attention was towards the capacity of the network, i.e., the amount of information that can be reliably transmitted in the network. While the proof of the validity of the approximation \cite{AvestimehrDT2011} is subtle, the intuition behind this work can be readily gleaned from a simple observation of the Cover-Wyner multiple access channel capacity region. Under high SNR regimes, the pentagon of the Cover-Wyner region can, in the limit, be decomposed into a rectangle, appended to a right isosceles triangle \cite{KimEYM2011}. The square can be interpreted as the communication region given by the bits that do not interfere. Such bits do not require special attention. In the case where the SNRs at the receiver for the different users are the same, this rectangle vanishes.
The triangular region is the same capacity region as for noise-free additive multiple access channel in a finite field \cite{EffrosM03}, leading naturally to an additive model over a finite field.

Note that, while we consider an equivalent additive finite-field additive model, this does not mean our underlying physical network  model is reliant on symbol-wise synchronization between the senders. Asynchrony among users does not affect the behavior of the underlying capacity region \cite{HH85}, on which the approximate model is predicated. Nor are users required to have the same received power in order to have the finite-filed equivalence hold -- differences in received power simply  lead to different shapes of the Cover-Wyner region, but the interpretation of the triangular and rectangular decomposition of the Cover-Wyner region is not affected. Moreover, our assumption of knowing the interfering users is fairly standard in multiple access wireless communications. Issues of synchronization, SNR determination and identification of users are in practice handled often jointly, since a signature serves for initial synchronization in acquiring the signal of a user, for measuring the received SNR and also for identification of the transmitting user. Finally note that, as long as we have appropriate coding, then the Cover-Wyner region  represents the region not only for coordinated transmissions, but also for uncoordinated packetized transmissions, such as exemplified in the classical ALOHA scheme \cite{Metal04}. This result, which may seem counterintuitive, is due in effect to the fact that the system will be readily shown to be stable as long as the individual and sum rates of the Cover-Wyner region will exceed the absolute value of the derivative of an appropriately defined Lyapunov function based on the queue length of a packetized ALOHA system.

There has been work on optimization of transmissions over the model of \cite{AvestimehrDT2011}. These approaches \cite{GoemansI09, AmaudruzF09, ShiR10} generally provide algorithms for code construction or for finding the maximum achievable rate, for multicast connections, over a high SNR network under the model of \cite{AvestimehrDT2011}. The approach of \cite{KimEYM2011, ErezX10, KimM10} considers a more general finite-field model and reduces the problem to an algebraic network coding problem \cite{KoetterM03}.  Random code constructions, inspired from \cite{HoM06} are then with high probability optimal. These approaches differ from our work in this paper in that they are interested in throughput maximization in a static model rather than completion delay when multiple transmission rounds may occur.
We are interested in  the latter model and, in particular, in how long it takes to broadcast  successfully a specific piece of information.

There are many deterministic and randomized algorithms for scheduling transmissions in wireless networks. They differ in some aspects of the model, such as whether the nodes can detect collision or cannot distinguish between a collision and silence, and whether the nodes know the entire network graph, or know only their neighbors, or do not have any such knowledge at all. Some papers that studied local broadcast are~\cite{KomlosG1985,CzyzowiczGKP2011}, where deterministic algorithms were presented, and~\cite{Martel94,JurdzinskiS2002,BienkowskiKKK2010}, which studied randomized algorithms.

In the setting of a wireless network, deterministic global broadcast of a single message was studied in~\cite{KowalskiP2005,DeMarco2010,CzumajR2003}, the best results given being $O(n\log{n})$ and $O(n\log^2{D})$, where $D$ is the diameter of the network. Bar-Yehuda et al.~\cite{BarYehudaGI87} were the first to study randomized global broadcast algorithms. Kowalski and Pelc~\cite{KowalskiP2005} and Czumaj and Rytter~\cite{CzumajR2003} presented randomized solutions based on selecting sequences, with complexities of $O(D\log{\frac{n}{D}}+\log^2{n})$. These algorithms match lower bounds of~\cite{KushilevitzM1993,AlonBNLP1991} but in a model that is weaker than the one addressed in this paper.
The algorithms mentioned above are all for global broadcast of one message from a known source. For multiple messages, a deterministic algorithm for $k$ messages with complexity $O(k\log^3{n}+n\log^4{n})$ appears in~\cite{ChlebusKPR2011},  while randomized global broadcast of multiple messages was studied in~\cite{Bar-yehuda93multiplecommunication,KKKL-DialM,KhabbazianKowalski-podc11}. We refer the reader to an excellent survey on broadcasting in radio networks in~\cite{Peleg 2007}.

Wireless networks are not always static; for example, nodes may fail, as a result of hardware or software malfunctions. Tolerating failed and recovered components is a basic challenge in decentralized systems because the changes to the network graph are not immediately known to nodes that are some distance away. Similarly, nodes may join and leave the network, or may simply be mobile. All of these cases result in changes to the network graph that affect communication. Depending on the assumptions, these changes can be quite arbitrary. Having a dynamic network graph imposes additional challenges on designing distributed algorithms for wireless networks. Dynamic networks have been studied in many papers. The problems addressed include determining the number of nodes in the network, gossiping messages, data aggregation, and distributed consensus~\cite{KuhnLO2010, CN10,KuhnOshman-disc11,KMO-podc11}. For global broadcast, some papers assume restrictions on the changes made in each round. For example,~\cite{ClementiMPS2009} consider graph changes that are random. They also consider the worst-case adversary, as do the studies in~\cite{KuhnLO2010,ODellW2005}.
In~\cite{KuhnLO2010} collisions are abstracted away, so that edges of the network graph do not represent nodes that hear the transmissions, but nodes that actually obtain the message. In~\cite{HaeuplerKarger}, the authors show how to use network coding in order to obtain more efficient algorithms for global broadcast in this dynamic model.

\section{Network Abstraction}
\label{sec:network}
We consider a wireless network where the transmission of a node is received at all neighboring nodes, perhaps colliding with transmissions of other nodes. Formally, the network is represented by an undirected graph $G=(V,E)$, where $|V|=n$. We denote by $N(u)$ the subset of $V$ consisting of all of $u$'s neighbors in $G$ and by $D$ the diameter of the network. The network topology is unknown.

We address two different radio models. One is the full-duplex model, in which nodes can listen to the channel while transmitting. The second is the half-duplex mode, in which at every time, a node can either transmit or listen to the channel. A transmission of a node $v\in V$ is modeled as a string of bits $\bar{s}_v$. The communication abstraction is such that the information received by a listening node $u \in V$ is equal to $\bigoplus_{v \in N(u)}\bar{s}_v$, where the operation $\oplus$ is the bit-wise XOR operation.  

The model is further assumed to be synchronous, that is, the nodes share a global clock, and a fixed slot length (typically $O(\polylog{n})$) is allocated for transmission.

Most of the paper assumes a bound $a \leq n$ on the contention that is known to all nodes. However, the actual contention in the network, which we denote by $a'$, may be even smaller than $a$. Each node has a unique ID from some set $I$ of size $|I|=N$, such that $N=n^{O(1)}$.

\section{Bounded-Contention Codes}
\label{sec:xor-codes}

To extract information from collisions, we propose the following coding technique for basic Bounded-Contention Coding, in which each node encodes its message into a codeword that it transmits, in such a way that a collision admits only a single possibility for the set of messages that produced it. This enables unique decoding.
\begin{definition}
\label{def:xor-code}
An $[M,m,a]$-BCC-code is a set $C \subseteq \{0,1\}^m$ of size $|C|=M$ such that for any two subsets $S_1,S_2 \subseteq C$ (with $S_1 \neq S_2)$ of sizes $|S_1|,|S_2| \leq a$ it holds that $\bigoplus{S_1} \neq \bigoplus{S_2}$, where $\bigoplus{X}$ of $X=\{\bar{x}_1,\dots,\bar{x}_{t}\}$ is the bit-wise XOR $\bar{x}_1\oplus\dots\oplus \bar{x}_{t}$.
\end{definition}

As a warm-up, we start by giving an example of a very simple BCC-code. This is the code of all unit vectors in $\{0,1\}^M$, i.e., $C=\{\bar{x}_i | 1\leq i \leq M, \bar{x}_i(j)=1 \text{ if and only if }i=j\}$. It is easy to see that $C$ is an $[M,M,M]$-BCC-code, since every subset $S\subseteq C$ is of size $s \leq M$, and we have $\bigoplus{S}=\bar{x}_s$, where $\bar{x}_s(j)=1$ if and only if $\bar{x}_i \in S$, implying that Definition~\ref{def:xor-code} holds.

The parameter $M$ will correspond to the number of distinct transmissions possible throughout the network; for example, it could correspond to the number of nodes. The parameters $a$ and $m$ will correspond to contention and slot length, respectively.
Therefore, the BCC-codes that interest us are those with $m$ as small as possible, in order to minimize the amount of communication. The above simple code, although tolerating the largest value of $a$ possible, has a very large codeword length. Hence, we are interested in finding BCC codes that trade off between $a$ and $m$. To show that such good codes exist, we need the following basic background on linear codes. An \emph{$[M,k,d]$-linear code} is a linear subspace of $\{0,1\}^M$, (any linear combination of codewords is also a codeword) of dimension $k$ and minimum Hamming weight $d$ (the Hamming weight of a codeword is the number of indexes in which the codeword differs from zero). The \emph{dual code} of a linear code $D$, denoted $D^{\bot}$, is the set of codewords that are orthogonal to all codewords of $D$, and is also a linear code. It holds that $(D^{\bot})^{\bot}=D$. The construction for arbitrary $[M,m,a]$-BCC-codes works as follows.

\textbf{BCC construction: }
Let $D$ be a linear code of words with length $M$ and Hamming weight at least $2a+1$. Let $\{\bar{x}_1,\dots,\bar{x}_m\}$ be a basis for the dual code $D^{\bot}$ of $D$. Let $H$ be the $m\times M$ parity-check matrix of $D$, i.e., the matrix whose rows are $\bar{x}_1,\dots,\bar{x}_m$, and let $C$ be the set of columns of $H$. We claim that $C$ is the desired BCC-code.
\begin{lemma}
\label{lemma:dual-code}
The code $C$ constructed above is an $[M,m,a]$-BCC-code.
\end{lemma}

\newcommand{\ProofLemmaDualCode}{
\begin{proofof}{Lemma~\ref{lemma:dual-code}}
It is clear from the construction that there are $M$ codewords in $C$, each of length $m$. Assume that $C$ is not an $[M,m,a]$-BCC-code. Then by Definition~\ref{def:xor-code}, there are two subsets $S_1,S_2 \subseteq C$ of sizes $|S_1|,|S_2| \leq a$, respectively, such that $\bigoplus{S_1}=\bigoplus{S_2}$. Without loss of generality we can assume that $S_1 \cap S_2 = \emptyset$ because adding the same codeword to both sets does not effect the equality of their XORs. This implies that $\bigoplus{S_1 \cup S_2}=0$. However, it is well known (see~\cite[Theorem 2.2]{Roth2006}) that every set of $d-1$ columns of $H$, where $d$ is the minimum Hamming weight of $D$, is linearly independent\footnote{For every set $S\subseteq C$, let $\bar{y}_S$ be the length $M$ vector characterizing the set $S$, i.e., $\bar{y}_S(i)=1$ if and only if the $i$-th column is an element in $S$. Consider the multiplication of the matrix $H$ by the vector $\bar{y}_S$. If $H\cdot \bar{y}_S^T=0$ then $\bar{y}$ is orthogonal to $D^{\bot}$, hence $\bar{y} \in (D^{\bot})^{\bot}=D$. This implies that the Hamming weight of $\bar{y}$ is at least $2a+1$.}. In our case, $d=2a+1$ and $|S_1 \cup S_2| \leq 2a$, which contradicts having $\bigoplus{S_1 \cup S_2}=0$, giving that $C$ is an $[M,m,a]$-BCC-code.
\end{proofof}
}

\ifabs{}{\ProofLemmaDualCode}
As the following sections will show, we need $[M,m,a]$-BCC-codes with $m$ as small as possible and $a$ as large as possible. By Lemma~\ref{lemma:dual-code}, this means we need to find linear codes of dimension $k=M-m$ as large as possible and minimum Hamming weight $d\geq 2a+1$ as large as possible. Note that we are only interested in the existence of good codes as the ID of a node will imply the codewords assigned to it, requiring no additional communication.

\begin{lemma}
\label{lemma:code-exist}
There is an $[M,m,a]$-BCC code with $m=O(a\log{M})$.
\end{lemma}

\newcommand{\ProofLemmaCodeExist}{
\begin{proofof}{Lemma~\ref{lemma:code-exist}}
The Gilbert-Varshamov bound~\cite{Roth2006} says that $2^k \geq \frac{2^M}{\sum_{j=0}^{d-1}{n \choose j}}$ for codes of length $M$, dimension $k$, and minimum Hamming distance $d$. This implies that $k \geq M - \log(d(\frac{Me}{d-1})^{d-1})$ which is $O(M-d\log{\frac{M}{d}})$. In our notation, this gives $m = O(a\log{M})$.
A greedy algorithm that repeatedly adds as a codeword an element of $\{0,1\}^M$ that is not in a ball of distance $d$ around any previously chosen codeword clearly attains the Gilbert-Varshamov bound. A slight modification also produces such a code that is also linear (see, e.g.,~\cite[Theorem 4.4]{Roth2006}). By Lemma~\ref{lemma:dual-code}, using this in the above BCC construction gives an $[M,m,a]$-BCC code with $m=O(a\log{M})$.
\end{proofof}
}
\ifabs{}{\ProofLemmaCodeExist}
Lemma~\ref{lemma:code-exist} implies, for example, that there are BCC-codes with $a=\Theta(\log{M})$ and $m=O(\log{M}\cdot\log{M})=O(\log^2{M})$. As explained earlier, for solving the problem of local broadcast, the parameters $a$ and $m$ correspond to the contention and the transmission length, respectively. As we show in the next section, such BCC-codes with polylogarithmic parameters are well-suited for the case of bounded contention, hence we refer to them as Bounded-Contention Codes.

In fact, the BCC-codes presented above are optimal, since $\Omega(a\log{M})$ is a lower bound for $m$. The reason for this is that each XOR needs to uniquely correspond to a subset of size at most $a$ out of a set of size $M$. The number of such subsets is $M \choose a$, therefore each codeword needs to have length $\Omega(\log{M \choose a})=\Omega(a\log{M})$.

%
%
\section{Local Broadcast}
\label{sec:local-bcast}
This section shows how to use BCC-codes for obtaining local broadcast in the additive radio network model. 
The simplest way to illustrate our technique is the following. Assume that in every neighborhood there are at most $a$ participants, and each node needs to learn the IDs of all participants in its neighborhood. The nodes use an $[N,a\log{N},a]$-BCC code to encode their IDs and, since at most $a$ nodes transmit in every neighborhood, every receiver is guaranteed to be able to decode the received XOR into the set of local participants. For the case of a single-hop network, we show 
how this information can then be used in order to assign unique transmission slots for the participants. However, while this shows the simplicity of using BCC-codes for coping with collisions, it does not extend to multi-hop networks since these require more coordination in order to assign slots for interfering transmitters (who can be more than a single hop from one another, as in the case of a \emph{hidden terminal}). Instead, we show how to use BCC-codes for directly coding the information rather than only the IDs of the transmitters.

\newcommand{\SingleHopSec}{
\subsection{Single-hop Networks}
\label{subsec:single-hop}
For the case of a single-hop network, the graph $G$ is a complete graph: in each time slot every listening node receives an XOR of all the strings transmitted in this slot by all other nodes in the network. We assume a bound $a$ on the contention, but our approach allows a slight improvement by being adaptive to the actual number of transmitters, $a'$, rather than only depending on its bound, $a$. We do this by assigning a few slots for the purpose of just finding out which are the nodes that want to transmit and agreeing on an allocation of slots for them in which each transmitter $v$ will transmit alone, guaranteeing that all other nodes receive its data $s_v$ successfully. We start with the full-duplex model, which means that nodes can listen and transmit at the same time.

\begin{algorithm}
1:\>Transmit codeword $C(u)$ from an $[N,a\log{N},a]$-BCC code $C$, \\
\>\>and receive $\bar{c}=\bigoplus_{v \in S}{C(v)}$, where $S\in V$ is the set of nodes that transmit.\\
2:\>Order nodes $S$ by increasing IDs and transmit $s_u$ at your turn.
\caption{Local broadcast using BCC, code for transmitter $u$ with data $s_u$.}
\label{alg:local-bcast-full}
\end{algorithm}


Since this is a single-hop network, all nodes receive $\bar{c}=\bigoplus_{v \in S}{C(v)}$, where $S\in V$ is the set of nodes that want to transmit. Recall that we are assuming only small contention in the network, which means formally that $|S| = a' \leq a$. Since $C$ is an $[N,a\log{N},a]$-BCC-code, $\bar{c}$ can be uniquely decoded into the set $\{C(v)|v\in S\}$, and, more important, the set $S$ can be uniquely identified by all nodes. This means that all nodes know after the first slot exactly who wants to transmit. This implies an agreement on the number of slots needed for each of the to transmit alone, as well as an order for their transmitters by increasing IDs.

After sending the codewords, the transmitters actually go ahead with their transmissions according to the slot assignment, which means that $a' \leq a$ slots are used, each of length $\ell$. This sums up to $O(a\log{n}+a'\ell)$ bits. We can improve this even further by noticing that we do not need to bound in advance the length $\ell$ of the information transmitted. The reason for this is that each transmitting node now has its own assignment, and could transmit for an unbounded number of slots while ending its transmission with some predetermined signal, allowing the next transmitter to go ahead. This does not require more than a very small constant overhead in the length of the transmission, and it implies adaptivity in terms of the message length (as well as in terms of the actual contention). This gives the following result.
\begin{theorem}
\label{theorem:encode-slots-singlehop-varmsglen} 
In a single-hop network with $a' \leq a$ transmitters $v_i$, $i \in \{1,\dots,a'\}$ with information of $\ell_{v_i}$ bits each, Algorithm~\ref{alg:local-bcast-full} gives that every node receives all the information within $O(a(\log{n})+\sum_{i=1}^{a'}{\ell_{v_i}})$ bits.
\end{theorem}

With half-duplex radios, we let each node choose whether it listens or transmits (if needed) with probability $1/2$. This gives that for every message and every node $v$, in each round there is probability $1/4$ for the message to be transmitted and heard by $v$. In expectation, a constant number of rounds is needed for $v$ to hear any single message, and using a standard Chernoff bound implies that $O(\log{n})$ rounds are needed with high probability. Finally, a union bound over all $n$ nodes and all messages gives the following theorem.
\begin{theorem}
\label{theorem:encode-info-singlehop-half-duplex}
In a single-hop network with $a' \leq a$ transmitters $v_i$, $i \in \{1,\dots,a'\}$ with information of $\ell_{v_i}$ bits each, the modification of Algorithm~\ref{alg:local-bcast-full} gives that every node receives all the information within $O(a(\log^2{n})+\sum_{i=1}^{a'}{\ell_{v_i}})$ bits, with high probability.
\end{theorem}
}
\ifabs{}{\SingleHopSec}

\ifabs{}
{
\subsection{Multi-hop Networks}
\label{subsec:multi-hop}

For multi-hop networks we first consider the same problem of local broadcast: every node, whether or not it is a transmitter, must receive all the messages sent by its neighbors. For this subsection, we assume that the bound $a$ on the contention is a \emph{local} bound, that is, there are at most $a$ nodes in every set $N(u)$ throughout the network that want to transmit.
}

We assume the real data that a node $v$ wants to transmit may be any element $s \in \{0,1\}^{\ell}$. Instead of encoding the IDs of the nodes in the network, we use an $[N2^{\ell},m,a]$-BCC code $C$ and every node $v$ is assigned $2^{\ell}$ codewords $\{C(v,s)| s \in \{0,1\}^{\ell}\}$ for it to use. This implies that the length of the codewords is $m=O(a\log{(N2^{\ell})})=O(a(\log{N}+\ell))=O(a(\log{n}+\ell))$. Notice that this is optimal since $a\log{n}$ is required in order to distinguish subsets of size $a$ among $n$ nodes, and the $a\ell$ term cannot be avoided if $a$ nodes transmit $\ell$ bits each. 

With half-duplex radios, we let each node choose whether it listens or transmits (if needed) with probability $1/2$. This gives that for every message and every node $v$, in each round there is probability $1/4$ for the message to be transmitted and heard by $v$. In expectation, a constant number of rounds is needed for $v$ to hear any single message, and using a standard Chernoff bound implies that $O(\log{n})$ rounds are needed with high probability. Finally, a union bound over all $n$ nodes and all messages gives the following theorem.

\begin{theorem}
\label{theorem:encode-local-multihop}
In a multi-hop network with at most $a$ transmitters with information of $\ell$ bits each in each $N(u)$, local broadcast can be obtained within $O((a\log{n}+a\ell)\log{n})$ bits, with high probability.
\end{theorem}


\section{Global Broadcast}
\label{sec:global-bcast}

In this section we show how to obtain global broadcast by combining BCC and network coding. We assume that at most $a$ nodes have a message of $\ell$ bits each that needs to be received by all nodes of the network. We first briefly introduce random linear network coding (RLNC) as a solution to the global broadcast problem in additive radio networks and then, in Subsection~\ref{subsec:RLNC+BCC}, show how BCC can significantly reduce the coding coefficient overhead of RLNC when $a << n$.

\subsection{Random Linear Network Coding}

RLNC is a powerful method to achieve optimal global broadcast, in particular in distributed networks in which nodes cannot easily coordinate to route information through the network. Instead of sending pieces of information around directly, RLNC communicates (random) linear combinations of messages over a finite field $F_q$. In this paper we will choose the field size $q$ to be $2$ which allows us to see vectors in $F_q$ simply as bit-vectors and linear combinations of vectors as XORs. 

We denote with $m_u \in F_2^{\ell}$ the message sent out by node $u$ and denote with $S$ the set of at most $a$ nodes that initially have a message. Given this, any packet sent out during the RLNC protocol has the form $(\mu,\sum_{u \in I} \mu_u m_u) \in F_2^{N+{\ell}}$ where $\mu \in F_2^{N}$ is a coefficient vector indicating which messages are XOR-ed together in the second portion of a packet, i.e., a characterizing vector. We call packets of this form \emph{valid}. A node $u$ that initially starts with a message $m_u$ treats this message as if it received the packet $(e_u,m_u)$ before round one, where $e_u$ is the standard basis vector corresponding to $u$ (that is, with a one at the coefficient corresponding to $u$, and zeros otherwise). During the protocol, each node that is supposed to send a packet takes all packets it has received so far and includes each of them independently with probability $1/2$ in the new packet. The new packet is formed by taking the XORs of all packets selected in this way (if no packet is selected the node remains silent or alternatively sends the all zero vector). Nodes decode by using Gaussian elimination. This can be done if and only if a node has received $a$ valid packets with linearly independent coefficient vectors. We note that, because of linearity, all initial packets and all packets created during the RLNC protocol are valid. More importantly, if multiple neighbors of a node send valid packets then the XOR of these packets which is received is also valid since the coefficient vectors and the message part XOR separately and component-wise. This makes RLNC a simple but powerful tool for exploiting the linear and additive nature of the additive radio networks we study in this paper.

\newcommand{\RLNC}
{
\begin{algorithm}
0:\> Initially, $S_u$ contains a single packet $(e_u,m_u)$\\
1:\> FOR round $i=1,\dots,32(D+a+\log{N})$:\\
2:\>\> With probability 1/2 DO:\\
3:\>\>\> Listen to channel for $N+\ell$ bits and update $S_u$,  the set of packets known to $u$\\
4:\>\> OTHERWISE\\
5:\>\>\> Send a packet $(\mu,\sum_{v \in S_u} \mu_v m_v)$,\\
6:\>\>\>\> which is an XOR of a subset of packets in $S_u$, each chosen with probability $1/2$.

\caption{Global broadcast using RLNC, code for transmitter $u$ with data $s_u$.}
\label{alg:RLNC}
\end{algorithm}
}
\ifabs{}{\RLNC}

We analyze the complexity of this RLNC scheme when used on top of an additive radio network. As in Section~\ref{sec:local-bcast}, nodes can either transmit or listen to the channel at any given round since they have half-duplex radios. We use the above RLNC algorithm together with the strategy of choosing in each round whether to transmit or listen at random with probability $1/2$.

We show that the RLNC protocol achieves an optimal round complexity of $O(D + a + \log n)$ with high probability. Our proof is based on the projection analysis technique from \cite{Haeupler2011} but we give a simple, self-contained proof \ifabs{in the full version of this paper.}{here.} The reason that the analysis carries over from a message passing model to the radio networks considered here so easily is their additivity. In particular, we use the effect that the XOR of randomly selected packets sent out by several neighbors which get XORed in the air are equivalent to the XOR of a random selection of packets known to at least one neighbor.

\begin{theorem}
\label{thm:standardRLNC}
RLNC disseminates all $a$ messages, with high probability, in $O(D + a + \log n)$ rounds in which messages of $N + {\ell}$ bits are sent in each round. 
\end{theorem}

\newcommand{\ProofThmStdRLNC}{
\begin{proofof}{Theorem~\ref{thm:standardRLNC}}
We say a node \emph{knows} a coefficient vector $\mu \in F_2^{N}$ if it has received a packet with a coefficient vector that is non-perpendicular to $\mu$ (over $GF(2)$). We claim that for any non-zero vector $\mu$ the probability that any fixed node $v$ does not learn $\mu$ within $O(D + a + \log n)$ rounds is at most $2^{-(a + 2 \log n)}$. Then, a union bound over all nodes and all $2^a$ coefficient vectors in $\mu \in F_2^{a}$ initially shows that, with high probability, all nodes know all vectors in the span of the messages given away initially. Finally, it is then easy to conclude that all nodes can decode. 

To prove this claim we look at a shortest path $P$ from $v$ to a node that initially knows $\mu$ (i.e., starts with a message with a non-zero coefficient in $\mu$). At any round $t$, let node $u$ be the closest node to $v$ on $P$ that knows $\mu$. There is a $1/2$ chance that $u$ sends and an independent $1/2$ chance of $1/2$ that $u$ sends out a packet with a coefficient vector that is non-perpendicular to $\mu$. Thus, in any round independently with probability at least $\frac{1}{4}$ knowledge of $\mu$ makes progress on $P$. Therefore, in $32(D + a + \log n)$ rounds the expected number of rounds that make progress is $8(D + a + \log n)$. A standard Chernoff bound shows that the probability that less than $D$ progress is made in these $32(D + a + \log n)$ rounds  is at most $2^{-(a + 2 \log n)}$ as claimed.
\end{proofof}
}

\ifabs{}{\ProofThmStdRLNC}
\subsection{Reducing the Overhead of Random Linear Network Coding via BCC-Codes}
\label{subsec:RLNC+BCC}

Note that \Cref{thm:standardRLNC} shows that RLNC has an essentially optimal round complexity. In particular, $\Omega(D)$ is a trivial lower bound since information passes at most one step in the network per round and $\Omega(a)$ is a lower bound too since in each round at most ${\ell}$ bits worth of information messages are received while $a{\ell}$ bits need to be learned in total. Lastly, the $\Omega(\log n)$ factor is tight for the proposed algorithm, too, because of the randomness used. On the other hand, the packets sent around have size $N + {\ell}$ while carrying only ${\ell}$ bits of information about the messages. Note that $N > n$ and in many cases $N = n^c >> {\ell}$ for some constant $c$ which renders the standard RLNC implementation highly inefficient. 

The reason for this is that we use an $N$ bit vector as a header to describe the set of IDs of nodes whose message is coded into the packet. This vector is always extremely sparse since at most $n<<N$ nodes are present and at most $a<<n$ nodes are sending a message. Instead of writing down a vector as is one could thus try to use a short representation of these sparse vectors. Writing down only the IDs of the non-zero components would be such a sparse representation (with almost optimal bit size $a \log N$) but does not work here, because when multiple neighbors of a node send sparse coefficient vectors their received XOR cannot be uniquely decoded. BCC-codes solve exactly this problem, by providing a \emph{sparse vector representation}:

\begin{definition}
Let $I$ be an ID set of size $N$ and $a$ be a sparseness parameter. Any $[N,a \log N, a]$-BCC code $C$ mapping any ID $u \in I$ to $C(u) \in F_2^{a \log N}$ induces a \emph{sparse vector representation} $s$ that maps the vector $\mu \in F_2^{N}$ to $s(\mu) = \sum_{u | \mu_u = 1} C(u)$.
\end{definition}

The following two properties make this representation so useful (in particular in this context):

\begin{lemma}
Let $I$ be an ID set of size N, $a$ be a sparseness parameter and $s$ be a sparse vector representation induced by a BCC-code $C$. For any two vectors $\mu, \mu' \in F_2^{N}$ with at most $a$ non-zero components we have:
\begin{itemize}
\item	Unique Decodability: $\ \mu \neq \mu'   \ \ \ \ \Longrightarrow \ \ \ \  s(\mu) \neq s(\mu')$.
\item	Homomorphism under addition:  $s(\mu) + s(\mu') = s(\mu + \mu')$.
\end{itemize}
\end{lemma}

Replacing the coefficient vectors $\mu$ in the RLNC scheme with their sparse representation leads to the much more efficient RLNC+BCC scheme.

As in the RLNC protocol, we denote with $m_u \in F_2^{\ell}$ the message sent out by node $u$ and denote with $S$ the set of at most $a$ nodes that initially have a message. Any packet sent out during the RLNC+BCC protocol has the form $(\mu,\sum_{u \in I} \mu_u m_u) \in F_2^{a\log{N}+{\ell}}$ where $\mu \in F_2^{a\log{N}}$ is a \emph{coded} coefficient vector indicating which messages are XOR-ed together in the second portion of a packet, i.e., it holds the XOR of the BCC-codewords of the IDs of the messages. As before, each node that is supposed to send a packet takes all packets it has received so far and includes each of them independently with probability $1/2$ in the new packet. The new packet is formed by taking the XORs of all packets selected in this way, preceded by the corresponding coded coefficient vector. Nodes decode by using Gaussian elimination, which can be done if and only if a node has received $a$ valid packets with linearly independent coefficient vectors. We note that, because of linearity, all initial packets and all packets created during the RLNC protocol are valid. Unlike having a list of IDs as a sparse representation of the coefficient vector, the power of BCC here is that if multiple neighbors of a node send valid packets then the XOR of these packets which is received is also valid since the BCC-coded coefficient vectors and the message part XOR separately and component-wise. Formally, the algorithm is identical to RLNC, except that the set $S_u$ of messages received by node $u$ is initialized to $(C(u),m_u)$ instead of $(e_u,m_u)$, and the node listens for $a\log{N}+\ell$ bits, rather than $N+\ell$.

\newcommand{\RLNCBCC}
{
\begin{algorithm}
0:\> Initially, $S_u$ contains a single packet $(C(u),m_u)$\\
1:\> FOR round $i=1,\dots,32(D+a+\log{N})$:\\
2:\>\> With probability 1/2 DO:\\
3:\>\>\> Listen to channel for $a\log{N}+\ell$ bits and update $S_u$,  the set of packets known to $u$\\
4:\>\> OTHERWISE\\
5:\>\>\> Send a packet $(\mu,\sum_{v \in S_u} \mu_v m_v)$,\\
6:\>\>\>\> which is an XOR of a subset of packets in $S_u$, each chosen with probability $1/2$.

\caption{Global broadcast using RLNC+BCC, code for transmitter $u$ with data $s_u$.}
\label{alg:RLNC-BCC}
\end{algorithm}
}
\ifabs{}{\RLNCBCC}

\begin{theorem}\label{thm:RLNC-BCC}
RLNC+BCC disseminates all $a$ messages, with high probability, in $O(D + a + \log n)$ rounds in which messages of $O(a \log n + {\ell})$ bits are sent in each round. 
\end{theorem}

\section{Dynamic Networks}
\label{sec:dyn}

In this section, we consider the case of a highly-dynamic network with a worst-case adversary: in every round, that is, between the times nodes send packets, the network graph is determined by the adversary, which observes the entire computation so far when deciding upon the graph for the next round. Notice that all of the above results for local broadcast hold for such dynamic networks, given that a slot length is sufficiently long in order to contain the required information. This is, for example, $O(a(\log{n}+\ell))$ bits in the full-duplex model, which is reasonable to assume if $a$ and $\ell$ are not too large. We get absolute guarantees for local broadcast in this highly dynamic setting, while existing work on avoiding collisions in radio networks cannot achieve this since they require probabilistic transmissions.

Next, we generalize the RLNC+BBC framework for the case of this highly-dynamic network with a worst-case adversary. The only restriction is that the graph has to be connected in every round. 
\ifabs{The proof of the resulting theorem is essentially the same as for the static case but instead of arguing that every message makes progress over a shortest path $P$, we argue that it makes some progress since the graph is always connected. Hence $D$ is replaced by $n$ in the number of rounds needed.}{The proof of the resulting theorem is essentially the same as for the static case but instead of arguing that every message makes progress over a shortest path $P$, we argue that it makes some progress since the graph is always connected. Hence $D$ is replaced by $n$ in the number of rounds needed.}

\begin{theorem}\label{thm:dynamicRLNC}
In a dynamic additive radio network controlled by an adaptive adversary subject to the constraint that the network is connected at every round, RLNC+BCC achieves global broadcast of $a$ messages, with high probability, in $O(n + a + \log n)$ rounds using a packet size of $O(a \log n + {\ell})$ bits.
\end{theorem}

\newcommand{\ProofThmDynRLNC}{
\begin{proofof}{Theorem~\ref{thm:dynamicRLNC}}
Let $S$ be a set of $a$ messages. We first analyze the algorithm given that $a$ is known, and then use the same estimation technique as before to address the case of an unknown $a$. For a coefficient vector $\mu \in F_2^{a}$ we measure progress by counting how many nodes know about it. Clearly, initially at least one node knows about $\mu$, while we want all $n$ nodes to know about it in the end. If in a round $r$ this is not achieved yet, we claim that there is a probability of at least $1/4$ for at least one more node to learn $\mu$. This is true because the graph is connected and hence there is at least one node $v$ that knows $\mu$ which is connected to a node $u$ that does not know $\mu$. As before, the node $u$ has a probability of at least $1/4$ to learn $\mu$ in this round $r$. As in the proof of \Cref{thm:standardRLNC}, a Chernoff bound shows that the probability that there were not enough (less than $n-1$) such successes in $32(n + a + \log n)$ rounds is at most $2^{-(a + \log n)}$. Finally, a union bound over all $2^a$ vectors completes the proof. For an unknown $a$, the same estimation technique works here since we only require progress through some path to every node, which is promised since the graph is connected at every round.
\end{proofof}
}

\ifabs{}{\ProofThmDynRLNC}

\section{Estimating the Contention}
\label{subsec:estimate}

We have given an almost optimal scheme for achieving global broadcast when the number of senders $a$ (or a good upper bound on it) is known a priori. This assumption is not an unreasonable one, for example, if network statistics show such behavior. However, in many cases, local contention may differ at different places throughout the network, or vary over time. It may also be that the known bound is pessimistic, and the actual contention is much smaller than this bound. In this section, we
show a method for removing this assumption by using BCC-codes to quickly determine $a$ (and also reveal the identity of all senders).

The mechanism we present allows for estimating the current contention and then using a code that corresponds to that estimate. A standard way to obtain a good estimation of contention is by having the nodes double a small initial guess until they succeed in local broadcast. In our BCC framework, the tricky part of this approach is to identify success. Specifically, using a bounded-contention code with parameter $a$ for a set $S$ of $k >a$ transmitters may produce an XOR that is a valid XOR of some set $S'$ of $k' \leq a$ transmitters. Hence, the nodes need to be able to distinguish between such a case and the case where $S'$ is the true set of transmitters.

The idea behind our algorithm is simple. We use an $[N,2k \log N,2k]$-BCC code to send out IDs in every round to make them propagate through the network. Every node $u$ keeps track of the set $S_u$ of all IDs it has heard from so far. In every round node $u$ sends out an XOR in which each of the IDs in $S_u$ is independently included with probability $1/2$. If $k \geq a$ then nodes receive the sum of at most $a$ nodes in every round and are able to split this sum into IDs which are then added to their sets. This way an ID propagates from one node to the next with constant probability and we show that within $O(D + \log n)$ rounds every node, with high probability, receives the ID of every node that wants to send. However, if $k<a$ we may get XORs of more than $2k$ IDs, which have no unique decoding guarantees by the BCC-code. The following algorithm takes care of this by detecting such a case eventually (and sufficiently fast). 


\begin{algorithm}
1: \> $k\leftarrow2$\\
2: \> REPEAT UNTIL $fail_u = false$ and $|S_u| \leq k$\\
3: \>\> $k\leftarrow 2k$\\
4: \>\> $fail_u \leftarrow false$\\
5: \>\> $C\leftarrow[N,2k \log N,2k]$-BCC code\\
6: \>\> IF node $u$ is a sender\\
7: \>\>\> $S_u \leftarrow \{C(u)\}$\\
8: \>\> ELSE\\
9: \>\>\> $S_u \leftarrow \emptyset$\\
10: \>\> FOR iteration $i=1,\dots,32(D + \log n)$:\\
11: \>\>\> IF $fail_u$\\
12: \>\>\>\> send $\log n$ random bits\\
13: \>\>\> ELSE \\
14: \>\>\>\> listen for $\log n$ random bits\\
15: \>\>\>\> IF received a non-zero string\\
16: \>\>\>\>\> $fail_u \leftarrow true$\\
17: \>\>\>With probability $1/2$ DO\\
18: \>\>\>\> Send $\sum_{v \in S_u} X_v C(v)$ where $X_v$ are i.i.d. uniformly Bernoulli \\
19: \>\>\>OTHERWISE\\
20: \>\>\>\> listen for $2k \log N$ bits\\
21: \>\>\>\> IF what received can be decoded as $\sum_{v \in S} C(v)$ for a $|S| \leq k$\\
22: \>\>\>\>\>  $S_u\leftarrow S_u \cup S$ \\
23: \>\>\>\> ELSE \\
24: \>\>\>\>\> $fail_u \leftarrow true$
 \caption{Estimating the Contention $a$, pseudocode for node $u$.} \label{algorithm:estimateContention}
\end{algorithm}

\begin{theorem}
\label{thm:estimateContention}
With high probability, Algorithm~\ref{algorithm:estimateContention} correctly identifies the subset of senders $S$ at every node after a total amount of communication of $O((D + \log n)(a \log n))$ bits.
\end{theorem}

\newcommand{\ProofThmEstimateContention}{
\begin{proofof}{Theorem~\ref{thm:estimateContention}}
We first show that at the end of each iteration in which $k<a$, with high probability, every node $u$ has $fail_u = true$ or $|S_u|>k$. To show this, we argue that for every node $u$ and every ID of a sender $s$, either $fail_u=true$ or $C(s) \in S_u$. Initially, by Lines 6-7, this is true for the sender $s$ itself. Let $P$ be a shortest path from $u$ to $s$, and let $w$ be the closest node to $u$ on $P$ that has $fail_w=true$ or $C(s) \in S_w$. If $fail_w=true$ then $w$ sends on Line 13, and this indication of failure makes progress on $P$, with high probability (the probability for a zeros string is exponentially small in its logarithmic length). Otherwise, $C(s) \in S_w$, and there is a probability of $1/2$ that $w$ transmits and an independent probability of $1/2$ that $w$ sends out a packet with $X_s = 1$ on Line 20.

Fix the set $X$ of IDs $x \neq s$ for which $C(x)$ is included in the XOR that $v$, the neighbor of $w$ along $P$, receives and consider the following cases. If $X$ is decoded into a set of size larger than $k$ or into a non-valid set then, with probability at least $1/4$, $C(s)$ is not included in the XOR and it is correctly decoded to produce $fail_v \leftarrow true$. Otherwise, $X$ is decoded into a set of size at most $k$ and then, with probability at least $1/4$, $C(s)$ is included in the XOR and $v$ decodes a valid set of size at most $k+1$. If the size is $k+1$ then $fail_v \leftarrow true $ and otherwise $C(s)$ is added to $S_v$.

Thus, in any round, independently with probability at least $1/4$, knowledge of $C(s)$ or the indication of failure makes progress on $P$. Therefore, in $32(D + \log n)$ rounds the expected number of rounds that make progress is $8(D + \log n)$. A standard Chernoff bound shows that the probability that less than $D$ progress is made in these $32(D + \log n)$ rounds  is at most $2^{-2\log n}$ as claimed. Then, a union bound over all $n$ nodes and all $k\leq n$ senders shows that, with high probability, every node $u$ has $fail_u = true$ or $S_u>k$. 

This implies that, with high probability, when $k<a$ all nodes double their estimate and proceed to the next iteration. The same analysis shows that if the algorithm reaches an iteration in which $k\geq a$ then the algorithm stops, with high probability, with each node having identified $S$, since a failure indication is never produced. The number of rounds in each iteration is $O(D + \log n)$ and the number of bits per round in an iteration with estimate $k$ is $O(k\log{N})$, which implies that the total number of bits communicated is $\sum_{k}{O((D+\log n)(k\log n))} = O((D + \log n)(a \log n))$.
\end{proofof}
}

\ifabs{}{\ProofThmEstimateContention}

One can use this procedure not just to estimate $a$ but also to exploit the fact that it gives the IDs of all senders in order to simplify the RLNC algorithm. For this, we order the IDs of the senders and assign to the $i$ highest node the $i$th standard basis vector out of the space $F_2^a$. We then use this as a sparse and concise coefficient vector in the RLNC protocol. This gives the following:

\begin{corollary}
After running the BCC-Estimation algorithm, RLNC achieves global broadcast, with high probability, in $O(D + a + \log n)$ rounds in which packets of size $a + \ell$ bits are sent in each round. 
\end{corollary}

\section{Discussion}

This paper presents a coding technique for additive wireless networks, which allows efficient local and global broadcast given a bound on the amount of contention. It also shows how to estimate the contention when it is not known in advance. The results hold also for dynamic networks whose arbitrary changes are controlled by a worst-case adversary. For full-duplex radios, it gives a deterministic framework providing absolute guarantees.

Directions for further research include using BCC-codes for solving additional distributed problems in the additive wireless network model, and handling extensions to the model, such as noise and asynchrony.

\small{
\paragraph{Acknowledgments.} \sloppypar The authors thank Seth Gilbert for useful discussions regarding probabilistic retransmissions, MinJi Kim and Ali ParandehGheibi for many discussions about the XOR collisions model addressed in this paper, and Amir Shpilka for pointing out the existence of the simple codes we use to implement our BCC framework. This work was supported in part by the Simons Postdoctoral Fellows Program and NSF Award 0939370-CCF.
}

\bibliographystyle{plain}	

\begin{thebibliography}{10}

\bibitem{AlonBNLP1991}
Noga Alon, Amotz Bar-Noy, Nathan Linial, and David Peleg.
\newblock A lower bound for radio broadcast.
\newblock {\em Journal of Computer and System Sciences}, 43:290--298, October
  1991.

\bibitem{AmaudruzF09}
Aurore Amaudruz and Christina Fragouli.
\newblock Combinatorial algorithms for wireless information flow.
\newblock In {\em SODA}, 2009.

\bibitem{AvestimehrDT2011}
Amir~Salman Avestimehr, Suhas~N. Diggavi, and David N.~C. Tse.
\newblock Wireless network information flow: A deterministic approach.
\newblock {\em IEEE Transactions on Information Theory}, 57(4):1872--1905,
  2011.

\bibitem{BarYehudaGI87}
Reuven Bar-Yehuda, Oded Goldreich, and Alon Itai.
\newblock On the time-complexity of broadcast in multi-hop radio networks: An
  exponential gap between determinism and randomization.
\newblock {\em J. Comput. Syst. Sci.}, 45(1):104--126, 1992.

\bibitem{Bar-yehuda93multiplecommunication}
Reuven Bar-Yehuda, Amos Israeli, and Alon Itai.
\newblock Multiple communication in multi-hop radio networks.
\newblock {\em SIAM Journal on Computing}, 22:875--887, 1993.

\bibitem{BienkowskiKKK2010}
Marcin Bienkowski, Marek Klonowski, Miroslaw Korzeniowski, and Dariusz~R.
  Kowalski.
\newblock Dynamic sharing of a multiple access channel.
\newblock In Jean-Yves Marion and Thomas Schwentick, editors, {\em 27th
  International Symposium on Theoretical Aspects of Computer Science (STACS
  2010)}, volume~5 of {\em Leibniz International Proceedings in Informatics
  (LIPIcs)}, pages 83--94, Dagstuhl, Germany, 2010. Schloss
  Dagstuhl--Leibniz-Zentrum fuer Informatik.

\bibitem{ChlebusKPR2011}
Bogdan~S. Chlebus, Dariusz~R. Kowalski, Andrzej Pelc, and Mariusz~A. Rokicki.
\newblock Efficient distributed communication in ad-hoc radio networks.
\newblock In Luca Aceto, Monika Henzinger, and Jiri Sgall, editors, {\em
  Automata, Languages, and Programming: 38th International Colloquium (ICALP
  2011, Part II), Zurich, Switzerland, July 2011}, volume 6755 of {\em Lecture
  Notes in Computer Science}, pages 613--624, 2011.

\bibitem{ClementiMPS2009}
Andrea E.~F. Clementi, Angelo Monti, Francesco Pasquale, and Riccardo
  Silvestri.
\newblock Broadcasting in dynamic radio networks.
\newblock {\em J. Comput. Syst. Sci.}, 75(4):213--230, 2009.

\bibitem{CN10}
Alejandro Cornejo and Calvin Newport.
\newblock Prioritized gossip in vehicular networks.
\newblock In {\em Proceedings of the 6th ACM SIGACT/SIGMOBILE International
  Workshop on Foundations of Mobile Computing (DIALM-POMC 2010)}, Cambridge,
  MA, September 2010.

\bibitem{CzumajR2003}
Artur Czumaj and Wojciech Rytter.
\newblock Broadcasting algorithms in radio networks with unknown topology.
\newblock {\em Proceedings of the 44th Annual IEEE Symposium on Foundations of
  Computer Science (FOCS 2003)}, page 492, 2003.

\bibitem{CzyzowiczGKP2011}
Jurek Czyzowicz, Leszek Gasieniec, Dariusz~R. Kowalski, and Andrzej Pelc.
\newblock Consensus and mutual exclusion in a multiple access channel.
\newblock {\em IEEE Transactions on Parallel and Distributed Systems},
  22(7):1092--1104, 2011.

\bibitem{DeMarco2010}
Gianluca De~Marco.
\newblock Distributed broadcast in unknown radio networks.
\newblock {\em SIAM Journal of Computing}, 39:2162--2175, March 2010.

\bibitem{EffrosM03}
M.~Effros, Muriel M\'{e}dard, T.~Ho, S.~Ray, D.~Karger, and R.~Koetter.
\newblock Linear network codes: A unified framework for source channel, and
  network coding.
\newblock In {\em Proceedings of the DIMACS workshop on network information
  theory (Invited paper)}, 2003.

\bibitem{ErezX10}
Elona Erez, Yu~Xu, and Edmund~M. Yeh.
\newblock Coding for the deterministic network model.
\newblock In {\em Allerton Conference on Communication, Control and Computing},
  2010.

\bibitem{GoemansI09}
Michel~X. Goemans, Satoru Iwata, and Rico Zenklusen.
\newblock An algorithmic framework for wireless information flow.
\newblock In {\em Proceedings of Allerton Conference on Communication, Control,
  and Computing}, 2009.

\bibitem{GollakotaK08}
Shyamnath Gollakota and Dina Katabi.
\newblock Zigzag decoding: combating hidden terminals in wireless networks.
\newblock In {\em SIGCOMM}, pages 159--170, 2008.

\bibitem{HaeuplerKarger}
Bernard Haeupler and David Karger.
\newblock Faster information dissemination in dynamic networks via network
  coding.
\newblock In {\em Proceedings of the 30th Annual ACM Symposium on Principles of
  Distributed Computing (PODC 2011)}, pages 381--390, San Jose, CA, June 2011.

\bibitem{Haeupler2011}
Bernhard Haeupler.
\newblock Analyzing network coding gossip made easy.
\newblock In {\em Proceedings of the 43rd annual ACM symposium on Theory of
  computing}, STOC '11, pages 293--302, New York, NY, USA, 2011. ACM.

\bibitem{HoM06}
Tracey Ho, Muriel M\'edard, Ralf Koetter, David~R. Karger, Michelle Effros, Jun
  Shi, and Ben Leong.
\newblock A random linear network coding approach to multicast.
\newblock {\em Information Theory, IEEE Transactions on}, 52(10):4413 --4430,
  2006.

\bibitem{HH85}
J.~Hui and P.~Humblet.
\newblock The capacity region of the totally asynchronous multiple-access
  channel.
\newblock {\em Information Theory, IEEE Transactions on}, 31(2):207 -- 216, mar
  1985.

\bibitem{JurdzinskiS2002}
Tomasz Jurdzinski and Grzegorz Stachowiak.
\newblock Probabilistic algorithms for the wakeup problem in single-hop radio
  networks.
\newblock In {\em Proceedings of the 13th International Symposium on Algorithms
  and Computation (ISAAC 2002)}, pages 535--549, London, UK, 2002.
  Springer-Verlag.

\bibitem{KhabbazianKowalski-podc11}
Majid Khabbazian and Dariusz Kowalski.
\newblock Time-efficient randomized multiple-message broadcast in radio
  networks.
\newblock In {\em Proceedings of the 30th Annual ACM Symposium on Principles of
  Distributed Computing (PODC 2011)}, San Jose, California, June 6-8 2011.

\bibitem{KKKL-DialM}
Majid Khabbazian, Dariusz Kowalski, Fabian Kuhn, and Nancy Lynch.
\newblock Decomposing broadcast algorithms using {A}bstract {MAC} layers.
\newblock In {\em Proceedings of Sixth ACM SIGACT/SIGMOBILE International
  Workshop on Foundations of Mobile Computing (DIALM-POMC 2010)}, Cambridge,
  MA, September 2010.

\bibitem{KimEYM2011}
MinJi Kim, Elona Erez, Edmund M., and Muriel M{\'e}dard.
\newblock Deterministic network model revisited: An algebraic network coding
  approach.
\newblock {\em CoRR}, abs/1103.0999, 2011.

\bibitem{KimM10}
Minji Kim and Muriel M\'edard.
\newblock Algebraic network coding approach to deterministic wireless relay
  network.
\newblock In {\em Allerton Conference on Communication, Control and Computing},
  2010.

\bibitem{KoetterM03}
Ralf Koetter and Muriel M\'edard.
\newblock An algebraic approach to network coding.
\newblock {\em IEEE/ACM Transaction on Networking}, 11:782--795, 2003.

\bibitem{KomlosG1985}
J{\'a}nos Koml{\'o}s and Albert~G. Greenberg.
\newblock An asymptotically fast nonadaptive algorithm for conflict resolution
  in multiple-access channels.
\newblock {\em IEEE Transactions on Information Theory}, 31(2):302--306, 1985.

\bibitem{KowalskiP2005}
Dariusz~R. Kowalski and Andrzej Pelc.
\newblock Broadcasting in undirected ad hoc radio networks.
\newblock {\em Distribed Computing}, 18:43--57, July 2005.

\bibitem{KuhnLO2010}
Fabian Kuhn, Nancy~A. Lynch, and Rotem Oshman.
\newblock Distributed computation in dynamic networks.
\newblock In {\em Proceedings of the 42nd ACM Symposium on Theory of Computing
  (STOC)}, pages 513--522, 2010.

\bibitem{KMO-podc11}
Fabian Kuhn, Yoram Moses, and Rotem Oshman.
\newblock Coordinated consensus in dynamic networks.
\newblock In {\em Proceedings of the 30th Annual ACM Symposium on Principles of
  Distributed Computing}, San Jose, California, June 6-8 2011.

\bibitem{KuhnOshman-disc11}
Fabian Kuhn and Rotem Oshman.
\newblock The complexity of data aggregation in directed networks.
\newblock In David Peleg, editor, {\em Distributed Computing: 25th
  International Symposium on Distributed Computing (DISC 2011), Rome, Italy,
  September 2011}, volume 6950 of {\em Lecture Notes in Computer Science},
  pages 416--431. Springer, 2011.

\bibitem{KushilevitzM1993}
Eyal Kushilevitz and Yishay Mansour.
\newblock An \protect{$\Omega(D\log(N/D))$} lower bound for broadcast in radio
  networks.
\newblock In {\em Proceedings of the 12th Annual ACM Symposium on Principles of
  Distributed Computing (PODC 1993)}, pages 65--74, New York, NY, USA, 1993.
  ACM.

\bibitem{Martel94}
Charles~U. Martel.
\newblock Maximum finding on a multiple access broadcast network.
\newblock {\em Information Processing Letters}, 52(1):7--15, 1994.

\bibitem{Metal04}
M.~Medard, Jianyi Huang, A.J. Goldsmith, S.P. Meyn, and T.P. Coleman.
\newblock Capacity of time-slotted aloha packetized multiple-access systems
  over the awgn channel.
\newblock {\em Wireless Communications, IEEE Transactions on}, 3(2):486 -- 499,
  march 2004.

\bibitem{ODellW2005}
Regina O'Dell and Roger Wattenhofer.
\newblock Information dissemination in highly dynamic graphs.
\newblock In {\em Proceedings of the 2005 joint workshop on Foundations of
  mobile computing}, DIALM-POMC '05, pages 104--110, New York, NY, USA, 2005.
  ACM.

\bibitem{ParandehGheibiS10}
Ali ParandehGheibi, Jay-Kumar Sundararajan, and M.~M\'edard.
\newblock Collision helps - algebraic collision recovery for wireless erasure
  networks.
\newblock In {\em WiNC}, 2010.

\bibitem{Peleg2007}
David Peleg.
\newblock Time-efficient broadcasting in radio networks: a review.
\newblock In {\em Proceedings of the 4th International Conference on
  Distributed Computing and Internet Technology (ICDCIT 2007)}, pages 1--18,
  Berlin, Heidelberg, 2007. Springer-Verlag.

\bibitem{Roth2006}
Ron~M. Roth.
\newblock {\em Introduction to coding theory}.
\newblock Cambridge University Press, 2006.

\bibitem{ShiR10}
Cuizhu Shi and Aditya Ramamoorthy.
\newblock Improved combinatorial algorithms for wireless information flow.
\newblock In {\em Proceedings of Allerton Conference on Communication, Control,
  and Computing}, 2010.

\end{thebibliography}

%
\end{document}